# Sparsity-based single-shot sub-wavelength coherent diffractive imaging

A. Szameit[1,4*], Y. Shechtman[1*], E. Osherovich[2*], E. Bullkich[1], P. Sidorenko[1], H. Dana[3], S. Steiner[4], E. B. Kley[4], S. Gazit[1], T. Cohen-Hyams[5], S. Shoham[3], M. Zibulevsky[2], I. Yavneh[2], Y. C. Eldar[6], O. Cohen[1], and M. Segev[1]

[1]*Physics Department and Solid State Institute, Technion, Haifa 32000, Israel*

[2]*Computer Science Department, Technion, 32000 Haifa, Israel*

[3]*Department of Biomedical Engineering, Technion, Haifa 32000, Israel*

[4]*Institute of Applied Physics, Friedrich-Schiller-Universität Jena, Max-Wien-Platz 1, 07743 Jena, Germany*

[5]*Microelectronics Research Center, Department of Electrical Engineering, Technion, Haifa 32000, Israel*

[6]*Department of Electrical Engineering, Technion, Haifa 32000, Israel*

\* These three authors contributed equally to this work

**Coherent Diffractive Imaging (CDI) is an algorithmic imaging technique where intricate features are reconstructed from measurements of the freely-diffracting intensity pattern [1-7]. An important goal of such lensless-imaging methods is to study the structure of molecules (including many proteins) that cannot be crystallized [8]. Clearly, high spatial resolution and very fast measurement are key features for many applications of CDI. Ideally, one would want to perform CDI at the highest possible spatial resolution and in a single-shot measurement - such that the techniques could be applied to imaging at ultrafast rates. Undoubtedly, such capabilities would give rise to unprecedented possibilities. For example, observing molecules while they dissociate or undergo chemical reactions will considerably expand the knowledge in physics, chemistry and biology. However, the resolution of all current CDI techniques is limited by the diffraction limit, and therefore**

cannot resolve features smaller than one half the wavelength of the illuminating light [9], which is considered a fundamental limit in diffractive imaging [10]. Moreover, combining CDI with current sub-wavelength imaging techniques would not allow for rapid single-shot measurements that are able to follow ultrafast dynamics, because such techniques rely on multiple exposures, either through mechanical scanning (e.g., Scanning Near-Field Microscope [11, 12], scanning a sub-wavelength "hot spot" [13-15]), or by using ensemble-averaging over multiple experiments with fluorescent particles [16, 17]. Here, we present sparsity-based single-shot sub-wavelength resolution in coherent diffraction microscopy: algorithmic reconstruction of sub-wavelength features from far-field intensity patterns of sparse optical objects. We experimentally demonstrate imaging of irregular and ordered arrangements of 100nm features with illumination wavelength of 532nm (green light), thereby obtaining resolutions several times better than the diffraction limit. The sparsity-based sub-wavelength imaging concept relies on minimization of the number of degrees of freedom, and operates on a single-shot basis [18-20]. Hence, it is suitable for capturing a series of ultrafast single-exposure images, and subsequently improving their resolution considerably beyond the diffraction limit. This work paves the way for ultrafast sub-wavelength CDI, via phase retrieval at the sub-wavelength scale. For example, sparsity-based methods could considerably improve the CDI resolution with x-ray free electron laser [21], without hardware modification. Conceptually, sparsity-based methods can enhance the resolution in all imaging systems, optical and non-optical.

Introduction

Improving the resolution in imaging and microscopy has been a driving force in natural sciences for centuries. Fundamentally, the propagation of an electromagnetic field in a linear medium can be fully described through the propagation of its eigen-modes (a complete and orthogonal set of functions which do not exchange power during propagation). In homogeneous, linear and isotropic media, the most convenient set of eigen-modes are simply plane-waves, each characterized by its spatial frequency and associated propagation constant. However, when light of wavelength $\lambda$ propagates in media with refractive index $n$, only spatial frequencies below $n/\lambda$ can propagate, whereas all frequencies above $n/\lambda$ are rendered evanescent and decay exponentially. Hence, for all propagation distances larger than $\lambda$, diffraction in a homogeneous medium acts as a low-pass filter. Consequently, optical features of sub-wavelength resolution appear highly blurred in a microscope, due to the loss of information carried by their high spatial-frequencies. Over the years, numerous "hardware" methods for sub-wavelength imaging have been demonstrated [11-17]; however, all of them rely on multiple exposures. Apart from hardware solutions, several algorithmic approaches for sub-wavelength imaging have been suggested (see, e.g. [22-24]). Basically, algorithmic sub-wavelength imaging aims to reconstruct the extended spatial frequency range (amplitudes and phases) of the information ("signal"), from measurements which are fundamentally limited to the range $[-n/\lambda, n/\lambda]$ in the plane-waves spectrum. However, as summarized in Goodman's 2005 book [25], "all methods for extrapolating bandwidth beyond the diffraction limit are known to be extremely sensitive to both noise in the measured data and the accuracy of the assumed a priori knowledge" such that "it is generally agreed that the Rayleigh diffraction limit represents a practical frontier that cannot be overcome with a conventional imaging system."

In spite of this commonly held opinion that algorithmic methods for sub-wavelength imaging are impractical [25], we have recently proposed a method for reconstructing sub-wavelength features from the far-field (and/or blurred images) of sparse optical information [18].

The concept of sparsity-based sub-wavelength imaging is related to Compressed Sensing (CS), which is a relatively new area in information processing [26-30]. We have shown that our sparsity-based method works for both coherent [18,20] and incoherent [19,31] light, and presented an experimental proof-of-concept [18,19]: the recovery of fine features that were cut off by a spatial low-pass filter. Subsequently, we took these concepts into the true sub-wavelength domain and demonstrated experimentally resolutions several times better than the diffraction limit: the recovery of 100nm features illuminated by 532nm wavelength light [20]. These ideas were followed by several groups, most notably the recent demonstration of sparsity-based super-resolution of biological specimens [32].

Here, we take the sparsity-based concepts into a new domain, and present the first experimental demonstration of sub-wavelength CDI: ***single-shot recovery of sub-wavelength images from far-field intensity measurements.*** That is, we demonstrate ***sparsity-based sub-wavelength imaging combined with phase-retrieval at the sub-wavelength level***. We recover the sub-wavelength features without measuring (or assuming) any phase information whatsoever; the only measured data at our disposal is the intensity of the diffraction pattern (Fourier power spectrum) and the support structure of the blurred image. Our processing scheme combines bandwidth extrapolation and phase retrieval, considerably departing from classical CS. We therefore devise a new sparsity-based algorithmic technique which facilitates robust sub-wavelength CDI under typical experimental conditions.

**Sparsity-based super-resolution**

In mathematical terms, the bandwidth extrapolation problem underlying sub-wavelength imaging corresponds to a non-invertible system of equations which has an infinite number of solutions, all producing the same (blurred) image carried by the propagating spatial frequencies. That is, after measuring the far field, one can add any information in the evanescent part of the spectrum while still being consistent with the measured image. Of course, only one choice corresponds to the correct sub-wavelength information that was cut off by the diffraction limit. The crucial task is therefore to extract the one correct solution out of the infinite number of possibilities for bandwidth extension. This is where sparsity comes into play. Sparsity presents us with prior information that can be exploited to resolve the ambiguity resulting from our partial measurements, and identify the correct bandwidth extrapolation which will yield the correct recovery of the sub-wavelength image.

Information is said to be sparse when most of its projections onto a complete set of base functions are zero (or negligibly small). For example, an optical image is sparse in the near-field when the number of non-zero pixels is small compared to the entire field of view. However, sparsity need not necessarily be in a near-field basis; rather, it can be in any mathematical basis. Many images are indeed sparse in an appropriate basis. In fact, this is the logic behind many popular image compression techniques, such as JPEG. In the fields of signal processing and coding theory, it is known for some time that a sparse signal can be precisely reconstructed from a subset of measurements in the Fourier domain, even if the sampling is carried out entirely in the low-frequency range [33]. This basic result was extended to the case of random sampling in the Fourier plane and initiated the area of CS [26]. An essential result of CS is that, in the absence of noise, if the "signal" (information to be recovered) is sparse in a basis that is sufficiently uncorrelated with the measurement basis, then searching for the sparsest solution (that conforms to the measurements) yields the correct solution. In the presence of noise (that is not too severe), the error is bounded, and many existing CS algorithms can recover the signal in a robust fashion under the same assumptions.

The concept underlying sparsity-based super-resolution imaging [18-20,31] and sparsity-based CDI relies on the advance knowledge that the optical object is sparse in a known basis. The concept yields a method for bandwidth extrapolation. Namely, sparsity makes it possible to identify the continuation of the truncated spatial spectrum that yields the correct image. As we have shown in [18], sparsity-based super-resolution imaging departs from standard CS, since the measurements are forced to be strictly in the low-pass regime, and therefore cannot be taken in a more stable fashion, as generally required by CS. Therefore, we developed a specialized algorithm, Non Local Hard Thresholding (NLHT), to reconstruct both amplitude and phase from low-frequency measurements [18]. However, NLHT, as well as other CS techniques necessitate the measurement of the phase in the spectral domain. In contrast, the current problem of sub-wavelength CDI combines phase-retrieval with sub-wavelength imaging, aiming to extrapolate the bandwidth from amplitude measurements only. Mathematically, this problem can be viewed in principle as a special case of quadratic CS, introduced in [31]. However, the algorithm suggested in [31] is designed for a more general problem resulting in high computational complexity. Here we devise a specified algorithm that directly treats the problem at hand.

**Sparsity-based sub-wavelength CDI**

For the current case of sub-wavelength CDI, the phase information in the spectral domain is not available. Hence, fundamentally, sub-wavelength CDI involves both bandwidth extrapolation and phase retrieval. However, despite the missing phase that carries extremely important information, we show that sparsity-based ideas can still make it possible to identify the correct extrapolation. Namely, if we *know* that our signal is sufficiently sparse in an appropriate basis, then - from all the possible solutions which could create the truncated spectrum - the correct extrapolation is often the one yielding the maximally sparse signal. Moreover, even under real experimental conditions, i.e., in the presence of noise, searching for the sparsest solution that is consistent with the measured data often yields a reconstruction that is very close to the ideal one.

Our algorithm iteratively reveals the support of the sought image by sequentially rejecting less likely areas (circles, in the experiments shown below). Thus, the sparsity of the reconstructed image increases with each iteration. This process continues as long as the reconstructed image yields a power-spectrum that remains in good agreement with the measurements. The process stops when the reconstructed power spectrum deviates from the measurements by some threshold value. However, it is important to emphasize that the exact threshold value and the degree of sparseness of the sought image need not be known a priori, as our method provides a natural termination criterion. Namely, *the correct reconstruction is identified automatically*. A detailed description of the reconstruction method, as well as comparison with other methods (that do not exploit sparsity), are provided in the Supplementary Information Section.

### Finding suitable basis

As explained above, sparsity-based CDI relies on the advance knowledge that the object is sparse in a known basis. In some cases, the 'optimal' basis – the basis in which the object is represented both well and sparsely - is known from physical arguments. For example, the features in Very Large Scale Integration (VLSI) chips are best described by pixels on a grid, because they obey certain design rules. In some cases, however, the prior knowledge about the optimal basis is more loose, namely, it may be known that the object is well and sparsely described in a basis that belongs to a certain family of bases. For example, one may know in advance that the object is sparse in the near field using a rectangular grid, yet the optimal grid spacing is not known a priory. We address this issue in section 4 of the Supplementary Information, where we describe a sparsity-based method that uses the experimental data to algorithmically find the optimal grid size (optimal basis) for our sub-wavelength CDI technique. That section also shows that the choice of basis functions is not particularly significant in our procedure: we obtain very reasonable reconstruction with almost any choice of basis functions, as long as they conform to the optimal grid. Finally, we note that recent work has shown that it is often possible to find the basis from a set of low-resolution images, using "blind CS" [34]. Likewise, in situations where a sufficient number of images of a similar type is available at high resolution, one can reconstruct the optimal basis through dictionary learning algorithms [35].

**Experiments**

We demonstrate sub-wavelength CDI technique experimentally on two-dimensional (2D) structures. The optical information is generated by passing a $\lambda$=532nm laser beam through an arrangement of nano-holes of diameter *100nm* each. The sample is made of a *100nm*-thick layer of chromium on glass; this thickness is larger than the skin depth at optical frequencies, such that the sample is opaque except for the holes. We use a custom microscope (NA=1, magnification x26) and a camera to obtain the blurred image. The optical Fourier transform of the optical information is obtained by translating the camera to the Fourier plane in the same microscope.

We begin with an ordered structure: a Star of David, consisting of 30 nano-holes. Figure 1a shows an SEM image of this sample. Figure 1b depicts the image seen in the microscope. As expected, the image is small and severely blurred. The spatial power spectrum (absolute value squared of the Fourier transform) of the image is shown in Fig. 1c. This truncated power spectrum covers a larger area on the camera detector, therefore facilitating a much higher number of meaningful measurements (each pixel corresponds to one measurement). We emphasize that only intensity measurements are used, in both the (blurred) image plane and in the (truncated) Fourier plane (Figs. 1b, 1c, respectively), without measuring (or assuming) the phase anywhere. The recovered image, using our sparsity-based algorithm, is shown in Fig. 1d. Clearly, we recover the correct number of circles, their positions, their amplitudes, and the entire spectrum (amplitude and phase), including the large evanescent part of the spectrum. *This demonstrates sub-wavelength Coherent Diffractive Imaging: image reconstruction combined with phase-retrieval at the sub-wavelength scale.* Moreover, as explained in the Supplementary Information section, the intensity of the blurred image (Fig. 1b) is used only for rough estimation of the image support. Our reconstruction method yields better results than other phase-retrieval algorithms (see comparisons in the Supplementary Information section), because it exploits the sparsity of the signal (the image to be recovered), as prior information. As mentioned earlier, the underlying logic is to minimize the number of degrees of freedom, while always conforming to the measured data, which in this case is the truncated power spectrum (intensity in Fourier space). In the example presented in Fig. 1, we take the data from Figs. 1b and 1c, search for the sparsest solution in the basis of circles of *100*nm diameter on a grid, and reconstruct a perfect Star of David, as shown in Fig. 1d. The grid is rectangular with *100*nm spacing (section 4 in the Supplementary Information describes how this parameter is found automatically), while the exact position of the grid with respect to the reconstructed information is unimportant (see Supplementary Information).

We emphasize that our reconstruction algorithm is able to reconstruct the phase in the spatial spectrum domain (the Fourier transform), from the intensity measurement in Fourier space and some rough estimation of the image support. In addition, we use the knowledge that the holes are illuminated by a plane wave, implying non-negativity of the image in real space. In this Star of David example, our algorithm reconstructs the phase in the spectral plane, as presented in Fig. 1e. For comparison, Fig. 1f shows the phase distribution in Fourier space, as obtained numerically from the ideal model of the subwavelength optical information (calculated from the SEM image of Fig. 1a). The reconstruction in Fig. 1 therefore constitutes the first demonstration of subwavelength CDI.

Interestingly, when comparing the Fourier transform of the sample with the measured spatial power spectrum, one finds that more than 90% of the power spectrum is truncated by the diffraction limit, acting as a low-pass filter (see Fig. 2). That is, we use the remaining 10% of the power spectrum and the blurred image, to successfully reconstruct the sub-wavelength features with high accuracy. In other words, the prior knowledge of sparsity and the basis is overcoming the loss of information in 90% of the power spectrum. As demonstrated in the Supplementary Information, it is the sparsity prior that makes it happen: without assuming the sparsity prior the reconstruction suffers from large errors.

The Star of David exhibits certain symmetries which could in principle assist the phase retrieval, had these symmetries been known. However, symmetry was not used for reconstruction of sub-wavelength features of Fig. 1. Nevertheless, it is illustrative to present another example with no spatial symmetry at all: an irregular arrangement of sub-wavelength holes on the assumed grid. Figure 3a shows the blurred image of an unknown number of sub-wavelength circles, distributed in a random manner. The respective Fourier power spectrum, as observed in the microscope, is shown in Fig. 3b. This sample is clearly not symmetric in real space, hence it does not exhibit a real Fourier transform. Still, we are able to reconstruct the sub-wavelength information, as shown in Fig. 3c, where all features of the original sample are retrieved, despite the inevitable noise in the experimental system. Figure 3d shows the SEM image of the sample, displaying the random arrangement of 100nm holes. The EM field passing through these nano-holes has roughly the same amplitude for all the holes. The reconstructed amplitudes at the hole sites are represented by the colors in Fig. 3c, highlighting the fact that the reconstructed field has similar amplitude at all the holes. The reconstructed phase in the spectral plane is presented in Fig. 3e, where the white circle marks the cutoff imposed by the diffraction limit. As shown there, our algorithm recovers the phase throughout the entire Fourier plane, including the region of evanescent waves far away from the cutoff frequency. For comparison, Fig. 3f shows the phase distribution in Fourier space, as obtained numerically from the ideal model of the sub-wavelength optical information (calculated from the SEM image of Fig. 3d). Clearly, the correspondence between the original spectral phase and the reconstructed one is excellent, including in the deep evanescent regions. Interestingly, Fig. 3e also displays the correct reconstruction of the phase around the faint high-frequency circle (of radius ~4 times the diffraction limit) where the phase jumps by $\pi$. Physically, this "phase-jump circle" is located at the first zero of the Fourier transform of a circular aperture, which in Fourier space multiplies the phase distribution generated by the irregular positions of the holes. The excellent agreement between Figs. 3e and 3f highlights the strength of the sparsity-based algorithmic technique.

**Discussion and Conclusions**

In this work, we presented a technique facilitating reconstruction of sub-wavelength features, along with phase-retrieval at the sub-wavelength scale, at an unprecedented resolution for single-shot experiments. That is, we have taken coherent lensless imaging into the sub-wavelength scale, and demonstrated sub-wavelength CDI from intensity measurements only. The method relies on prior knowledge – that the sample is sparse in a known basis (circles on a grid, in the examples in Figs. 1-3). We emphasize that *sparsity* is what makes our phase-retrieval work: the other assumptions used in the algorithm (non-negativity, bounded support and the known basis) alone are not sufficient. It is important to note that most natural and artificial objects are sparse, in some basis. The information does not necessarily have to be sparse in real space: it can be sparse in any mathematical basis whose relation to the measurement basis is known, e.g., the wavelet basis or the gradient of the field intensity, given that this basis is sufficiently uncorrelated with the measurements. In all these cases our technique can provide a major improvement by "looking beyond the resolution limit" in a single-shot experiment. Since our approach is purely algorithmic, it can be applied to every optical microscope and imaging system as a simple computerized image processing tool, delivering results in real time with practically no additional hardware. The fact that our technique works in a single-shot holds the promise for ultrafast sub-wavelength imaging: one could capture a series of ultrafast blurred images, and then off-line processing will reveal their sub-wavelength features, which could vary from one frame to the next. Finally, we note that our technique is general, and can be extended also to other, non-optical, microscopes, such as atomic force microscope, scanning-tunnelling microscope, magnetic microscopes, and other imaging systems. We believe that the microscopy technique presented here holds the promise to revolutionize the world of microscopy with just minor adjustments to current technology: sparse sub-wavelength images could be recovered by making efficient use of their available degrees of freedom. Last but not least, we emphasize that our approach is more general than the particular subject of optical sub-wavelength imaging. It is in fact a universal scheme for recovering information beyond the cut-off of the response function of a general system, relying only on a priori knowledge that the information is sparse in a known basis. As an exciting example, we have recently investigated the ability to utilize this method for recovering the actual shape of very short optical pulses measured by a slow detector [36]. Our preliminary theoretical and experimental results indicate, unequivocally, that our method offers an improvement by orders of magnitude beyond the most sophisticated deconvolution methods. In a similar vein, we believe that our method can be applied for spectral analysis, offering a means to recover the fine details of atomic lines, as long as they are sparse (i.e., do not form bands). In principle, the ideas described here can be generalized to any sensing / detection / data acquisition schemes, provided only that the information is sparse in a known basis, and that the measurements are taken in a basis sufficiently uncorrelated to it.

# Figure Captions

**Caption Fig. 1:**

**Reconstruction of two-dimensional sub-wavelength information**

(a) SEM image of the sample. (b) The blurred image, as seen in the microscope. The individual holes cannot be resolved. (c) The measured spatial power spectrum of the field. The color map is identical to that in the other panels: only the background was removed for clarity. (d) The reconstructed 2D information, algorithmically recovered from the measured power spectrum (Panel (c)) and the blurred image (Panel (b)). (e) The algorithmically recovered phase in the spatial spectrum domain. White represents zero-phase, while black denotes π-phase. (f) The phase in the spectral domain calculated from the Fourier transform of Panel (a).

**Caption Fig. 2:**

**Truncation of the information by the transfer function**

The absolute value of the Fourier transform of the field forming the Star of David. The spectrum is truncated by the diffraction limit (the white circle, corresponding to the zoom-in on the right), beyond which all spectral components are evanescent. The truncated part, containing more than 90% of the spectrum, is the measured power spectrum displayed in Fig. 1c.

**Caption Fig. 3:**

**Reconstruction of an irregular arrangement of two-dimensional sub-wavelength holes**

(a) The blurred image as seen in the microscope. (b) The measured spatial power spectrum of the field. (c) The algorithmically reconstructed 2D information, showing 12 holes in an irregular arrangement. (d) SEM image of the sample depicting the irregular arrangement of nano-holes. (e) The algorithmically recovered phase in the spatial spectrum domain. (f) The phase in the spatial spectrum domain calculated from the Fourier transform of Panel (d).

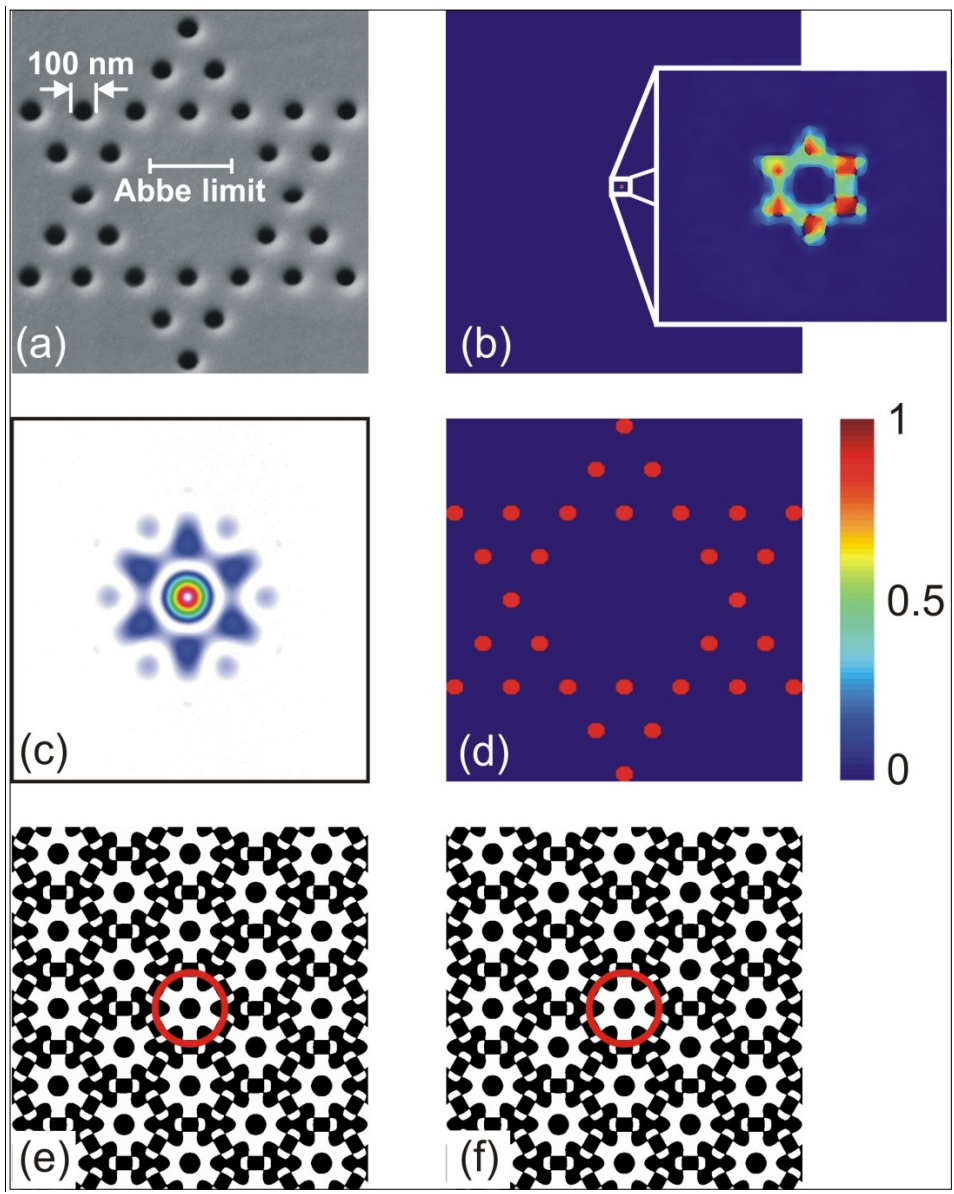

**Fig. 1**

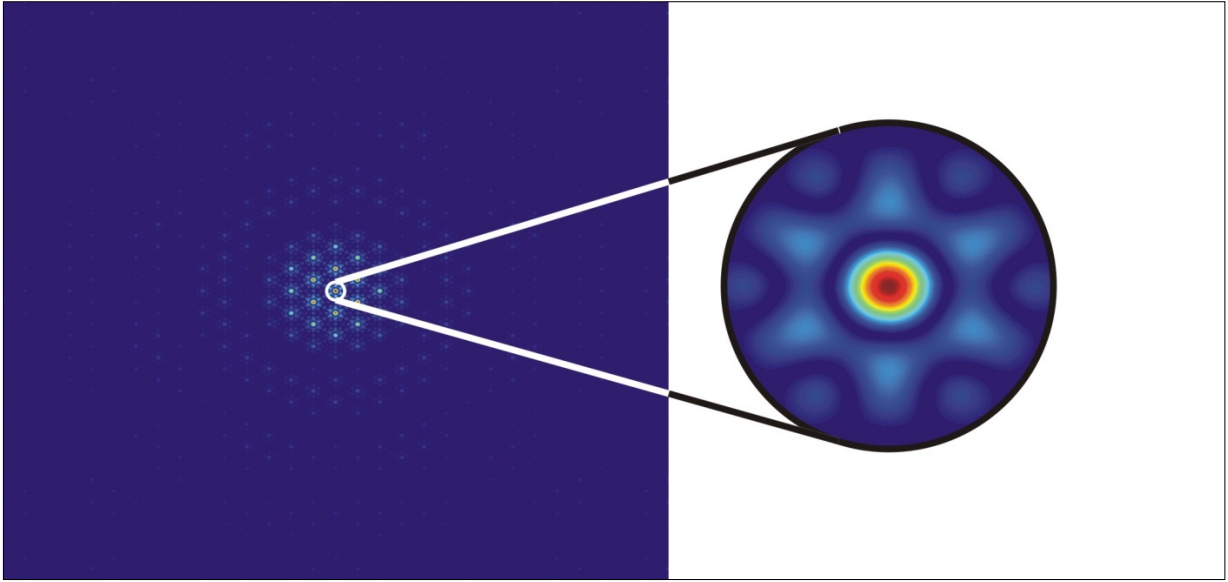

**Fig. 2**

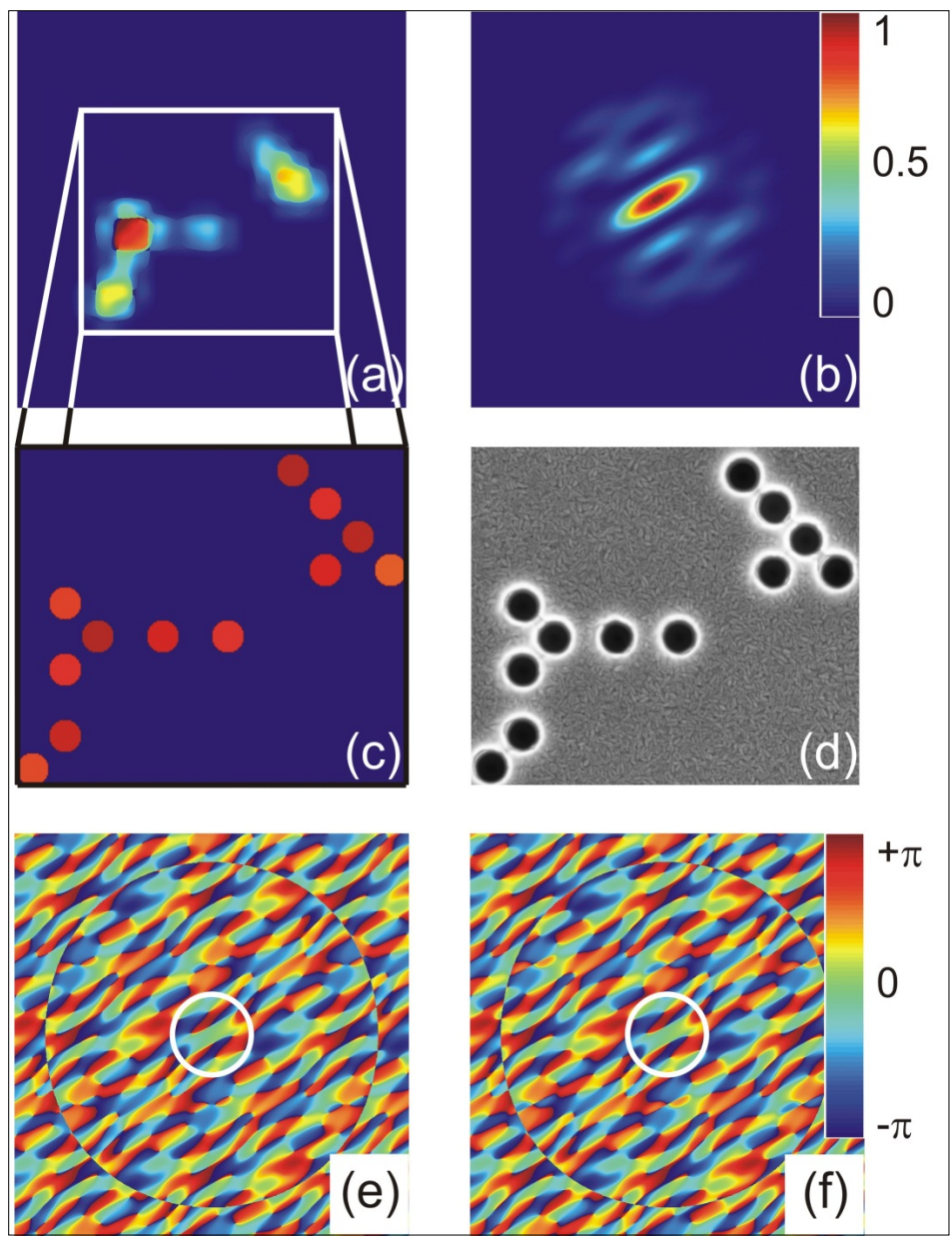

**Fig. 3**